\documentclass[amsmath,superscriptaddress,reprint,aps]{revtex4-2}
\usepackage{graphicx}
\usepackage{dcolumn}
\usepackage{bm}
\usepackage{hyperref}
\usepackage{siunitx}
\usepackage{color, soul, ulem}
\setstcolor{red}

\begin{document}
	
	
	\title{Quantitative cavity-enhanced photothermal dynamics in TMDC-integrated ultrahigh-Q microcavities}

      \author{Hidetoshi~Kanzawa}
      	\affiliation{Department of Physics, Faculty of Science and Technology, Keio University, Yokohama 223-8522, Japan}

         \author{Ryo~Sugano}
	\affiliation{Department of Electronics and Electrical Engineering, Faculty of Science and Technology, Keio University, Yokohama 223-8522, Japan}
 
     \author{Hajime~Kumazaki}
	\affiliation{Department of Physics, Faculty of Science and Technology, Keio University, Yokohama 223-8522, Japan}
    
         \author{Yuta~Takahashi}
	\affiliation{Department of Physics, Faculty of Science and Technology, Keio University, Yokohama 223-8522, Japan}

    \author{Mayori~Terada}
	\affiliation{Department of Physics, Faculty of Science and Technology, Keio University, Yokohama 223-8522, Japan}

         \author{Shinichi~Watanabe}
	\affiliation{Department of Physics, Faculty of Science and Technology, Keio University, Yokohama 223-8522, Japan}
    
	\author{Shun~Fujii}
	\email[Corresponding author. ]{shun.fujii@phys.keio.ac.jp}
 	\affiliation{Department of Physics, Faculty of Science and Technology, Keio University, Yokohama 223-8522, Japan}
	
	
\begin{abstract}
We investigate photothermal effects in monolayer transition metal dichalcogenides (TMDCs) integrated with an ultrahigh-Q silica microcavity. Launching a continuous-wave laser into a cavity resonance enables controlled intracavity heating, allowing direct observation of excitonic photoluminescence (PL) modulation.  A distinct redshift of the PL peak energy is observed as the pump wavelength is tuned across resonance. This behavior is quantitatively reproduced by a temperature-dependent bandgap model that combines the Varshni relation with the thermo-optic response of the microcavity, from which the local temperature rise can be estimated. We further find that PL collected through a fiber waveguide exhibits spectral and temporal characteristics markedly different from free-space emission, indicating selective coupling of the microcavity to specific excitonic channels.  These results provide a quantitative framework for understanding photothermal effects in TMDC-microcavity hybrid systems and offer a versatile approach for all-optical control and probing of thermal states in integrated nanophotonic devices.

\end{abstract}

	\maketitle
	

\section{Introduction}
Ultrahigh-quality (ultrahigh-Q) optical microcavities are widely used in diverse areas, including sensing~\cite{Cao2024, Liu2021}, nonlinear optics~\cite{lin2017, kippenberg2004}, optomechanics~\cite{aspelmeyer2014, Barzanjeh2022}, and quantum optics~\cite{strekalov2016}. Their small mode volumes and strong intracavity field enhancement enable efficient light-matter interactions, which also lead to increased optical absorption within the cavity material. This absorption inevitably induces temperature changes in the microcavity, modifying both the refractive index and physical dimensions through thermo-optic and thermal expansion effects, and consequently shifting the resonance frequencies~\cite{Carmon2004}. These photothermal effects play a crucial role in governing the dynamical response of optical microcavities and provide an efficient mechanism for tuning cavity resonances. Such thermally induced resonance shifts have been widely exploited in applications such as resonance control~\cite{Heylman2013} and thermal sensing~\cite{Liao2021, Dong2009, Zhu2014, Watts2007}.

In recent years, increasing attention has been paid to the use of optical microcavities to harness the unique optical properties of atomically thin two-dimensional (2D) van der Waals materials. When thinned down to the monolayer limit, transition metal dichalcogenides (TMDCs) exhibit a direct bandgap and broken inversion symmetry, giving rise to remarkable optical properties such as strong photoluminescence (PL)~\cite{Mak2010} and nonlinear optical responses~\cite{You2018}. Graphene and hexagonal boron nitride (hBN) also display unique properties that have not been exploited in conventional bulk materials. Owing to these characteristics, the integration of 2D materials with optical microcavities provides an attractive platform for exploring optical and optoelectronic applications, including lasing~\cite{Wu2015, Salehzadeh2015, Zhao2018, Fu2020, Ye2015}, excitonic emission~\cite{Reed2015, Reed2016, Khelifa2020, JaverzacGaly2018}, nonlinear frequency conversion~\cite{Fryett2018, Fujii2024, Liu2025}, high-repetition-rate mode-locked lasers~\cite{Kovalchuk2024}, and efficient single-photon emitters~\cite{Parto2022, Yang2025a}.

In such hybrid devices, strong light-matter interaction leads to enhanced photothermal effects in 2D materials integrated on the surface of the microcavity. In particular, optical absorption can induce significant local heating and has been reported to cause irreversible degradation in TMDC-based devices~\cite{Reed2016}. These observations highlight the need for precise characterization of the thermal state in microcavity-integrated 2D materials. At the same time, the enhanced photothermal response can be exploited to induce thermo-optic resonance shifts~\cite{Gao2017, Yuan2018, Jiang2025}, enabling various device functionalities such as photodetectors, optical switches, and modulators. Moreover, photothermal effects on the excitonic properties of integrated TMDCs remain insufficiently understood.
A quantitative understanding of local heating in such hybrid systems is therefore essential for both reliable device operation and optimal design.

Here, we investigate dynamic photothermal effects and their impact on excitonic photoluminescence in monolayer TMDCs integrated with an ultrahigh-Q whispering gallery mode (WGM) microcavity. By coupling a continuous-wave (CW) pump laser to a high-Q cavity mode, we induce controlled photothermal heating and directly monitor the resulting spectral evolution of excitonic emission from WSe$_2$ and WS$_2$. We observe a distinct redshift of the PL peak energy under pump wavelength tuning, which is quantitatively explained by a temperature-dependent bandgap model based on the thermo-optic response of the cavity. In addition, we show that the emission collected through the fiber waveguide exhibits distinct spectral and temporal characteristics, including a pronounced redshift and modified decay dynamics. These observations provide a comprehensive picture of cavity-enhanced photothermal effects and emission-channel selectivity in TMDC-microcavity hybrid systems, offering important insights for the design of 2D material-integrated nanophotonic devices, as well as for all-optical sensing of the thermal state of optical microcavities.

\section{Device fabrication, characterization, and thermal simulations}

\begin{figure*}
	\centering
		\includegraphics[width=1\textwidth]{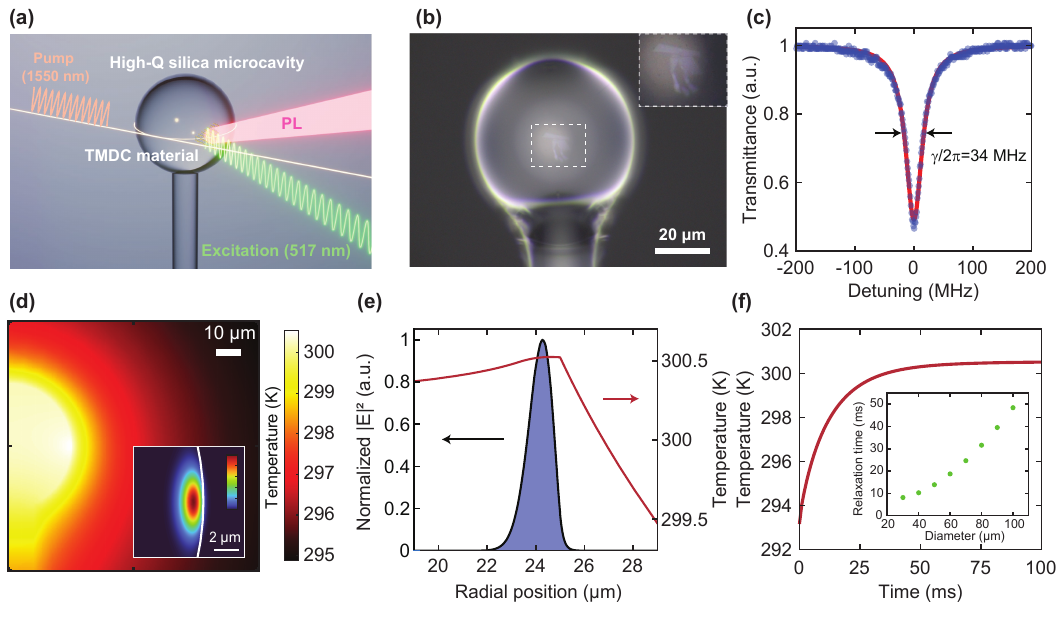}
		\caption{\label{Fig_concept} (a) Conceptual illustration of a TMDC-integrated ultrahigh-Q silica microsphere. (b) Optical micrograph of a silica microsphere decorated with a WSe$_2$ monolayer. (c) Typical transmission spectrum of the microcavity after deposition of the WSe$_2$ monolayer, yielding a Q factor of $\sim6\times10^6$. (d) Finite-element simulation of heat diffusion in a silica microsphere with a diameter of 50~\textmu m. The inset shows the optical mode profile. An optical input power of 10~mW is assumed in the thermal simulation. (e) Simulated normalized electric field amplitude of the optical mode and the corresponding temperature profile. (f) Simulated temporal response of the cavity-mode temperature induced by a localized heat source. The inset shows the thermal relaxation time as a function of the cavity diameter.}
\end{figure*}

Figure~\ref{Fig_concept}(a) shows a schematic illustration of the TMDC-integrated microsphere cavity of interest in this work. A silica microsphere with a diameter of approximately 70~\textmu m is fabricated by melting the tip of an optical fiber (SMF-28, Corning) using a commercial fiber splicer (31S, Fujikura)~\cite{Fujii2024}. The sphere diameter can be controlled by tapering the fiber prior to melting. This preprocessing step allows the fabrication of microspheres with diameters of less than 100~\textmu m. Monolayer TMDCs ($\mathrm{WSe_2}$ and $\mathrm{WS_2}$) are mechanically exfoliated from bulk crystals (HQ Graphene) onto a polydimethylsiloxane (PDMS) sheet (Gel-Pak), and the layer number is initially identified either through PL spectroscopy or by optical contrast in microscope images prior to dry transfer. The exfoliated monolayer flakes are then transferred onto the equator of the silica microsphere to maximize the interaction between the cavity modes and the TMDCs. A micrograph of the fabricated device is shown in Fig.~\ref{Fig_concept}(b).

The Q-factor of the fabricated microcavity is first characterized by coupling a wavelength-tunable CW laser (TSL-710, Santec) into the cavity via a tapered fiber. We obtain a typical Q-factor value of $6\times10^6$ by monitoring the resonance linewidth in the transmission spectrum at a wavelength of $\sim$1550~nm (Fig.~\ref{Fig_concept}(c)). The loaded Q-factor is estimated from the full width at half maximum (FWHM) of the resonance, $\gamma/2\pi$, where $\gamma=\omega/Q$ and $\omega/2\pi$ is the resonance frequency. The frequency is calibrated using a Mach-Zehnder interferometer~\cite{Fujii2020}. The polarization of the pump laser is adjusted using a fiber polarization controller, and the wavelength is monitored with a high-precision wavelength meter (86122C, Keysight).

The Q-factors of TMDC-integrated microspheres are primarily limited by additional scattering losses, as a pristine cavity before transfer exhibits ultrahigh-Q values of up to $\sim7\times10^7$. Such a degradation of the Q-factor by about one order of magnitude is consistent with previous studies~\cite{JaverzacGaly2018,Fujii2024,Liu2025}, and in turn indicates strong interaction between the optical mode and the surface-integrated monolayer flake.

Thermal effects induced by optical absorption cannot be neglected in optical microcavities because the small cavity dimensions significantly enhance temperature rise as well as heat conduction. As the thermal response time is proportional to the heat capacity divided by the thermal conductivity~\cite{Carmon2004}, microcavities exhibit faster thermal time constants, typically ranging from several tens of nanoseconds~\cite{Shakespeare2024} to several milliseconds~\cite{Chen2016,Zhou2021,Yang2025}. The thermal time constant reflects the response speed of photothermal conversion and is therefore a key parameter for device applications. We perform finite-element-method (FEM) simulations (COMSOL Multiphysics) to study the photo-induced thermal dynamics of a silica microsphere. The optical cavity mode (eigenstates of a microsphere cavity) is obtained by solving the Helmholtz equation, and its intensity distribution is used as the spatial profile of the heat source.

Figure~\ref{Fig_concept}(d) shows the steady-state temperature distribution of a silica microsphere with a diameter of 50~\textmu m, and the inset shows the normalized profile of the fundamental optical mode. The simulation indicates that efficient thermal diffusion in the silica microsphere leads to a relatively uniform temperature distribution despite the small optical mode area. Figure~\ref{Fig_concept}(e) presents the simulated electric field distribution and temperature distribution along the equatorial plane of the microsphere. The radius of the microsphere is 25~\textmu m, at which the TMDC monolayer is assumed to be deposited. Notably, the region of elevated temperature strongly overlaps with the optical mode maximum near the microsphere surface, indicating efficient local heating at the TMDC-integrated interface. The temperature difference between the TMDC-integrated surface and the cavity-mode region is therefore negligible on the scale of the cavity temperature rise considered in our analysis. Although possible excitation of higher-order modes may introduce uncertainty in the local heat-source distribution, the FEM analysis also indicates that heat diffusion leads to an approximately uniform temperature profile near the TMDC-integrated surface (Supplementary~1). Figure~\ref{Fig_concept}(f) shows the simulated temporal response of the cavity-mode temperature, and the inset shows the dependence of the thermal relaxation time on the cavity diameter.

\begin{figure*}
	\centering
		\includegraphics[width=1\textwidth]{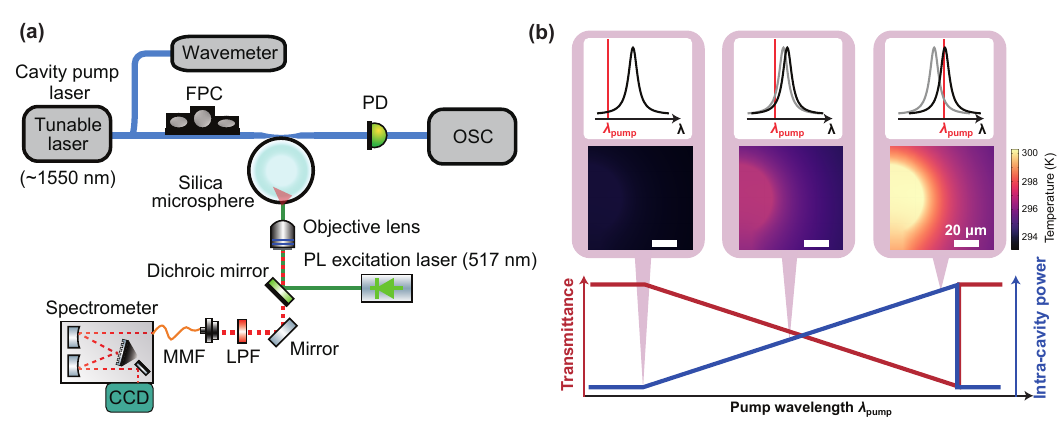}
		\caption{\label{Fig_setup} 
    (a) Photoluminescence (PL) measurement setup under microcavity excitation using a 1550~nm CW laser. FPC, fiber polarization controller; PD, photodetector; OSC, oscilloscope; LPF, long-pass filter; MMF, multimode fiber; CCD, charge-coupled device. (b) Schematic illustration of the thermal self-stabilization regime during continuous laser wavelength scanning. The laser wavelength is swept from shorter to longer wavelengths to exploit the thermally induced redshift of the cavity resonance. As the laser wavelength approaches the resonance, enhanced coupling increases the intracavity heating and reduces the cavity transmission. When the system exits the thermally self-stabilized regime, the transmission and intracavity power rapidly return to their initial values.
        }
\end{figure*}

\section{Experimental setup and thermal self-stabilization in optical microcavities}

To investigate the effect of optically induced temperature changes on monolayer TMDCs, we measure the PL spectrum while an external CW laser (pump laser) at $\sim$1550~nm is coupled to a high-Q cavity mode. The experimental setup is shown in Fig.~\ref{Fig_setup}(a), where PL excitation is provided by a 517~nm laser diode (GSL52A, Thorlabs), operated in CW mode. The PL emission is collected and detected by a low-noise charge-coupled device (CCD) detector (PIXIS-256E, Princeton Instruments) mounted on a spectrometer (HRS-300, Princeton Instruments). The excitation power for PL measurements is kept at $\sim$50~\textmu W to minimize possible photo-induced perturbations of the monolayer, including sample degradation, exciton–exciton annihilation, carrier-density-dependent spectral shifts, and local heating. The excitation laser beam is focused near the top region of the microsphere to maintain stable and reproducible optical alignment. The diameter of the focused spot is approximately 0.8~\textmu m. 

For cavity-mode excitation, the pump laser wavelength is gradually tuned toward a cavity resonance from the blue-detuned side, to ensure thermal self-stabilization~\cite{Carmon2004}. The pump power is 10~mW, and the polarization is adjusted to match the cavity mode using a polarization controller. The transmitted light is monitored with an oscilloscope after photodetection to evaluate the coupling condition. Schematics of thermal self-stabilization and the evolution of the transmittance, intracavity power, and simulated temperature distribution are shown in Fig.~\ref{Fig_setup}(b). The continuous decrease in transmittance accompanying the laser wavelength tuning provides direct evidence of thermal self-stabilization, where thermal effects stabilize the laser frequency at a certain laser-cavity detuning through a balance between absorption-induced heating and the thermally induced resonance shift. This mechanism counteracts laser frequency drift and effectively locks the laser to the cavity resonance without any active feedback control~\cite{Carmon2005}. It should be noted that this mechanism is key to sensing~\cite{Dong2009, Zhu2014, Watts2007}, thermal resonance locking~\cite{Carmon2005, Mcrae2009}, and access to nonlinear oscillation regimes~\cite{Zhang2019,Fujii2024}.

\section{Photothermal modulation of PL emission in cavity-integrated monolayer TMDCs}

\begin{figure*}
	\centering
		\includegraphics[width=1\textwidth]{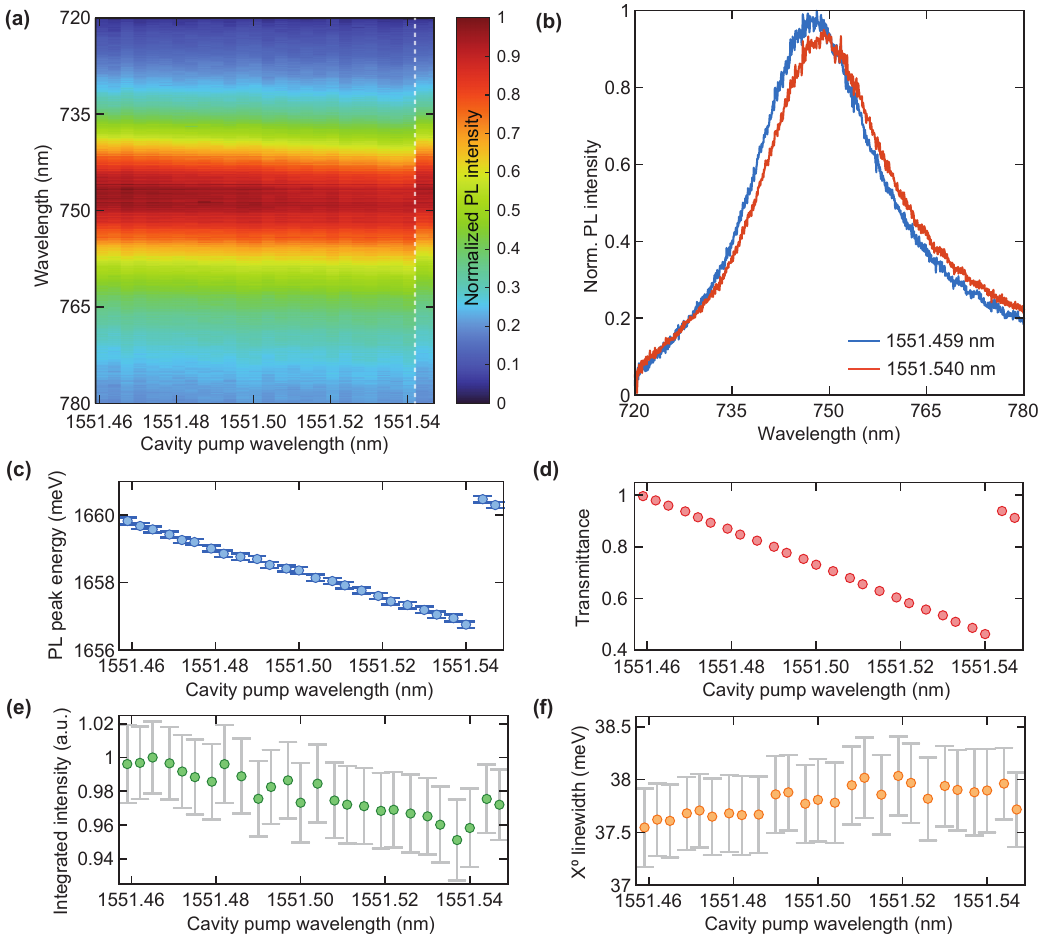}
		\caption{\label{Fig_PL} 
    (a) Normalized PL spectra of microcavity-integrated monolayer WSe$_2$ under thermal self-stabilization tuning. The pump wavelength is swept by $\sim$0.88~\AA\ from the blue- to red-detuned side. (b) PL spectra of WSe$_2$ at the initial (blue) and final (orange) stages of the self-stabilization process, respectively. The observed redshift is $\sim$3.0~meV. (c) Exciton peak energy extracted from the PL spectra. (d) Normalized cavity transmittance during the cavity pump wavelength sweep. (e) Normalized integrated intensity of the exciton emission. (f) Fitted linewidth of the exciton emission. The PL spectra are fitted using a double-Gaussian function to extract the exciton peak energy, integrated intensity, and linewidth.}
\end{figure*}

The continuous evolution of the PL spectrum is tracked during the laser scanning, as shown in Fig.~\ref{Fig_PL}(a). We measure two cavity devices decorated with monolayer WSe$_2$ and WS$_2$. The results presented in Fig.~\ref{Fig_PL} correspond to the WSe$_2$ device, while the measurements for WS$_2$ are provided in Supplement~1. As the pump laser wavelength is carefully tuned toward a cavity resonance, the PL peaks exhibit a redshift of a few meV, which is clearly observed in both the PL maps and the extracted spectra (Figs.~\ref{Fig_PL}(a) and \ref{Fig_PL}(b)). The evolution of the exciton peak energy obtained from spectral fitting is summarized in Fig.~\ref{Fig_PL}(c).

This PL redshift is accompanied by a distinct decrease in the pump transmittance, as shown in Fig.~\ref{Fig_PL}(d), indicating the onset of thermal self-stabilization in the microcavity. As the intracavity power increases near resonance, absorption-induced heating leads to a thermally induced redshift of the cavity mode, which in turn stabilizes the laser-cavity detuning. In addition to the energy shift, the integrated PL intensity decreases by approximately 5\% (Fig.~\ref{Fig_PL}(e)), and the exciton ($X^0$) linewidth broadens by approximately 2\% (Fig.~\ref{Fig_PL}(f)). These changes are consistent with local heating effects, where increased temperature reduces the PL intensity~\cite{Huang2016} and enhances exciton-phonon scattering, leading to linewidth broadening~\cite{Selig2016}.
The abrupt change in the PL spectrum observed at $\sim$1551.54~nm corresponds to the point at which the thermal self-stabilization is lifted, indicating the release from the thermally locked state. 

Here, we attempt to explain the observed PL redshift in terms of photo-induced thermal dynamics and temperature-dependent bandgap changes in TMDC monolayers. The dynamical thermal behavior of microcavity resonances, which gives rise to thermal bistability, has been well established both theoretically and experimentally~\cite{Carmon2004,Zhou2021,Jiang2020}. A coupled-mode simulation including the thermo-optic resonance shift is also performed to quantitatively evaluate the relationship among cavity transmission, intracavity power, and temperature rise during thermal self-stabilization (Supplementary~1). The resonance wavelength of a microsphere cavity is influenced by both the thermo-optic effect and thermal expansion. The thermo-optic effect refers to the temperature-dependent change in the refractive index $dn/dT$, whereas thermal expansion describes the change in the resonator geometry $dl/dT$. These effects modify the resonance condition, given by $2\pi n(T) r(T) = m\lambda(T)$, where $r$ is the cavity radius and $m$ is the azimuthal mode number, leading to a spectral shift of the cavity resonance. Since both $dn/dT$ and $dl/dT$ are positive for fused silica, the resonance wavelength exhibits a redshift as the cavity temperature increases. Assuming thermal bistability arising from these effects, the variation in the resonance wavelength can be written as:
\begin{equation}\label{Eq_ResonanceShift} 
\lambda(\Delta T) = \lambda_0 + \lambda_0
\left(
\frac{1}{n}\frac{dn}{dT}
+ \frac{1}{l}\frac{dl}{dT}
\right)\Delta T
= \lambda_0(1 + \zeta \Delta T),
\end{equation}
where $\Delta T$ is the temperature difference between the cavity mode and its surroundings, $\lambda_0$ is the cold (i.e., without pumping) cavity resonance, $n$ is the refractive index, and $\zeta$ is a coefficient that includes both thermo-optic and thermal expansion effects. The coefficient $\zeta$ is calculated to be $6 \times 10^{-6}~\mathrm{^\circ C^{-1}}$ for silica microcavities~\cite{Carmon2004}, providing a direct relation between the cavity pump wavelength and the temperature increase.  This relation enables us to quantitatively link the cavity-induced temperature rise to the observed shift in the PL energy.

With regard to the thermal effect on TMDC monolayers, the temperature dependence of the bandgap energy is described by the empirical Varshni equation~\cite{Varshni1967}:
\begin{equation}\label{Eq_Varshni}
E_g(T) = E_g(0) - \frac{\alpha T^2}{\beta + T},
\end{equation}
where $E_g(0)$ is the extrapolated bandgap energy at $0~\mathrm{K}$.

Equations~\eqref{Eq_ResonanceShift} and \eqref{Eq_Varshni} establish a direct relationship among three physical quantities: the cavity pump wavelength, the effective temperature of the cavity mode, and the PL peak energy. Through this relationship, the bandgap modification of the TMDC induced by the photothermal effect can be quantitatively estimated from the experimentally accessible pump laser wavelength. To verify this mechanism, the observed redshift of the PL peak energy is compared with the bandgap shift estimated from the corresponding temperature change. By directly comparing the measured PL energy shifts with the theoretically estimated bandgap variation, we evaluate the influence of cavity-induced heating on the TMDC monolayer. This comparison provides strong evidence that the observed PL redshift originates from photothermal effects in the microcavity.

\begin{figure*}
	\centering
		\includegraphics[width=1\textwidth]{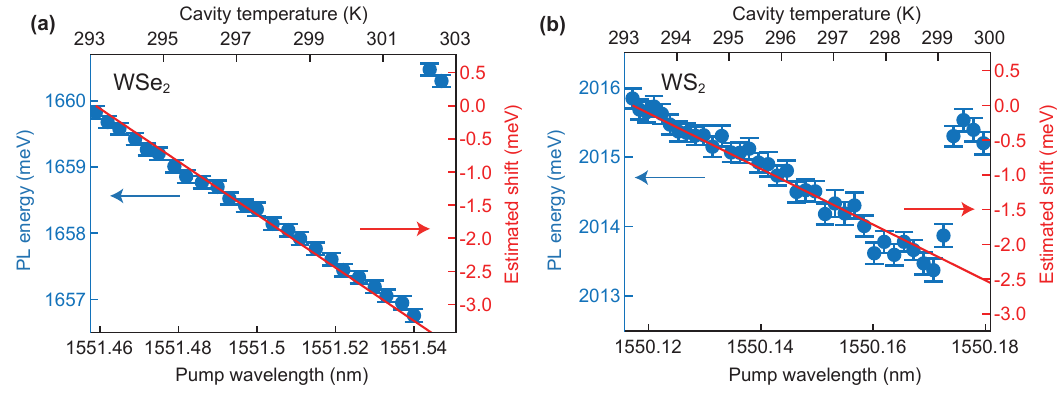}
		\caption{\label{Fig_peakshift} 
        Observed PL energy shift as a function of the pump wavelength and the corresponding cavity temperature calculated from Eq.~(\ref{Eq_ResonanceShift}) (blue circles), together with the estimated bandgap energy change (red lines) based on Varshni's empirical law (Eq.~(\ref{Eq_Varshni})). The two slopes show good agreement, indicating that the observed PL redshift originates from photothermal heating of the monolayer induced by cavity-enhanced light absorption. Results for WSe$_2$ (a) and WS$_2$ (b), respectively.
        }
\end{figure*}

The observed PL peak energy shifts of WSe$_2$ and WS$_2$ under pump wavelength tuning are plotted in Figs.~\ref{Fig_peakshift}(a) and \ref{Fig_peakshift}(b), respectively, as a function of the cavity pump wavelength. Using Eq.~\eqref{Eq_ResonanceShift}, the pump wavelength axis is converted into an effective cavity-mode temperature, allowing us to relate the measured PL energy shift to the corresponding temperature rise. Because the pump wavelength is scanned slowly compared with the simulated thermal relaxation time of the microsphere, the PL redshift observed during the scan can be treated as a quasi-steady-state photothermal response. Assuming that the pump wavelength remains close to the cavity resonance, the cavity-mode temperature is estimated to increase from the ambient temperature of 293.15~K to approximately 302~K during the tuning process. 

The red lines in Figs.~\ref{Fig_peakshift}(a) and \ref{Fig_peakshift}(b) represent the PL energy shifts estimated from the temperature change using the Varshni relation given in Eq.~\eqref{Eq_Varshni}, with coefficients reported by Kopaczek \textit{et al.}~\cite{Kopaczek2022} ($\alpha= 0.44$~meV/K ($0.45$~meV/K) and $\beta=190$~K (210~K) for $\mathrm{WSe_2}$ $(\mathrm{WS_2})$). It should be noted that the coefficients show good agreement with those obtained in our separate experiments (Supplementary~1). The left vertical axis shows the measured PL energy shift, while the right vertical axis represents the corresponding temperature-induced energy shift estimated from cavity heating, allowing direct comparison between experiment and theory. The initial temperature is assumed to be room temperature (293.15~K). For the WSe$_2$ device, tuning the pump wavelength from 1551.453~nm to 1551.540~nm results in a PL redshift of approximately 1.43~nm, corresponding to an energy shift of approximately 3.2~meV. For the WS$_2$ device, tuning the pump wavelength from 1550.117~nm to 1550.171~nm results in a PL redshift of approximately 0.75~nm, corresponding to an energy shift of approximately 2.5~meV. The 95\% confidence intervals obtained from the spectral fitting are included in Fig.~\ref{Fig_peakshift}.
Further tuning toward longer wavelengths releases the system from the thermally self-stabilized state, restoring the initial condition. 
It should be noted that the pump wavelength tuning range is determined primarily by cavity-resonance properties, such as the Q-factor, coupling condition, and polarization matching, rather than by the TMDC material itself. The measured PL energy shifts show good agreement with those estimated from the cavity-mode temperature increase, supporting the interpretation that the observed PL redshift originates from cavity-enhanced photothermal heating and subsequent heat transfer to the monolayer. These results suggest that PL emission from TMDC monolayers can serve as a sensitive probe of the thermal state of high-Q microcavities.

\section{PL emission properties from free-space and through a tapered fiber waveguide}

\begin{figure*}
	\centering
		\includegraphics[width=1\textwidth]{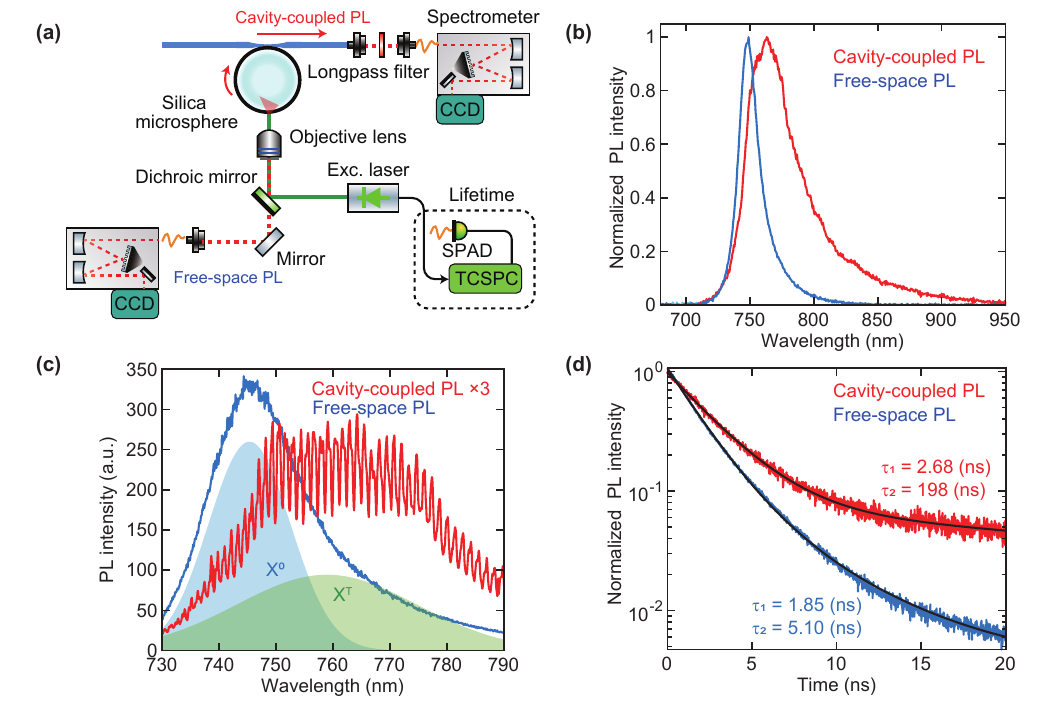}
		\caption{\label{Fig_decay}
        (a) PL spectroscopy setup combining free-space PL collection with cavity-coupled PL collection through a fiber waveguide. No 1550~nm continuous-wave (CW) excitation is applied in this configuration. The PL lifetime is evaluated using a time-correlated single-photon counting (TCSPC) system. SPAD, single-photon avalanche diode.(b) Overview of the PL spectra of monolayer WSe$_2$ acquired directly in free space (blue) and via the fiber waveguide (red). A pronounced spectral redshift is observed in the fiber-coupled spectrum. (c) Corresponding PL spectra measured with higher wavelength resolution ($\sim0.1$~nm). The free-space PL spectrum is fitted with two Gaussian components, attributed to the neutral exciton $X^0$ and the charged exciton $X^\mathrm{T}$. (d) Time-resolved PL intensity of the cavity-coupled emission extracted through the tapered fiber (red) and the free-space radiative PL emission (blue). The black lines represent biexponential fits.
        }
\end{figure*}

To further investigate the emission properties of monolayer TMDCs coupled to the microcavity, and to clarify the role of different collection pathways, we perform an additional PL measurement without cavity pumping at 1550~nm, as shown in Fig.~\ref{Fig_decay}(a). In hybrid systems of high-Q microcavities and nanomaterials, not only free-space emission but also waveguide-coupled emission is essential for practical applications. In particular, highly-efficient coupling to a waveguide enables selective access to cavity-coupled emission channels that possess several advantages such as enhanced collection efficiency and spectral filtering in photonic integrated platforms~\cite{JaverzacGaly2018,Yamashita2021,He2021,Parto2022,Tonndorf2017}.
Here, the microcavity is not optically pumped, and the PL emission is collected through two independent optical pathways: one is directly acquired in free space, while the other is collected via a tapered fiber that captures the emission coupled to the microcavity through evanescent coupling. 

The PL spectra obtained from these two pathways are shown in Fig.~\ref{Fig_decay}(b), while higher-resolution measurements over a narrower wavelength range are presented in Fig.~\ref{Fig_decay}(c), revealing a clear distinction between the two emission pathways. Interestingly, the free-space PL exhibits a central wavelength of approximately 745~nm, whereas the cavity-coupled PL shows a pronounced redshift with a central wavelength around 763~nm. The spectral spikes observed in the cavity-coupled PL originate from the discrete cavity modes, and the overall PL lineshape is significantly broadened compared to the free-space PL. The periodic spacing of these spikes corresponds to the free spectral range (FSR) of the silica microsphere, approximately 1.8~nm. Although discrete spectral spikes corresponding to WGM resonances are observed, the cavity-coupled PL spectrum retains a broad envelope because the excitonic emission couples to multiple closely spaced cavity modes. In addition, coupling to multiple transverse and higher-order WGM modes may further increase the density of resonant features, resulting in a broad cavity-coupled PL spectrum with superimposed WGM resonances. We note that a similar spectral redshift has been observed in similar configurations~\cite{Fujii2024, JaverzacGaly2018, Khelifa2020}, although the mechanism remains unclear.

To clarify the origin of the pronounced redshift in the cavity-coupled PL, we independently measure the broadband transmission spectrum of the tapered fiber alone. No wavelength-dependent transmission feature capable of reproducing the observed redshift is found, excluding passive spectral filtering by the taper-cavity system as the dominant origin (Supplementary~1). Fitting the free-space PL spectrum with a double-Gaussian function reveals that the long-wavelength peak, attributed to charged excitons (trions, $X^T$), is located close to the spectral position of the cavity-coupled PL (Fig.~\ref{Fig_decay}(c)). This observation could suggest that radiative emission originating from non-negligible out-of-plane (OP) dipole transitions (i.e., long-lived dark states), whose contribution is typically weak in conventional free-space collection, is enhanced through coupling to the characteristic propagation modes of the WGM microcavity as reported in Ref.\cite{AndresPenares2021}. In addition, reabsorption by the monolayer WSe$_2$ may further contribute to the attenuation of emission near 750~nm~\cite{Khelifa2020}, corresponding to the bandgap energy.

Finally, we perform time-resolved PL measurements to further investigate this unique phenomenon. The PL emission is detected using a single-photon avalanche diode (SPAD), and the photon histogram is recorded via time-correlated single-photon counting (TCSPC). The excitation laser is operated in a pulsed mode. For the free-space PL emission, the decay dynamics are well described by a biexponential function with lifetimes of $\tau_1 = 1.85$~ns and $\tau_2 = 5.10$~ns, as shown in Fig.~\ref{Fig_decay}(d). In contrast, the PL emission coupled from the cavity modes into the tapered fiber exhibits significantly longer decay components of $\tau_1 = 2.68$~ns and $\tau_2 = 198$~ns.  Although the emitted photons are coupled to the WGMs, the cavity-coupled PL exhibits a longer-lived decay component than the free-space PL. This suggests that the cavity-coupled emission is not influenced by the Purcell effect, which would typically lead to lifetime shortening~\cite{Fong2026}. This result is qualitatively consistent with the spectral characteristics discussed in the context of the cavity-enhanced OP dipole transition, where the cavity-coupled PL appears at longer wavelengths than the free-space PL.

The pronounced redshift and extended lifetime observed in the cavity-coupled emission indicate that the tapered-fiber collection does not simply sample the same excitonic population as free-space detection. Instead, the microcavity acts as a wavelength- and mode-selective channel that preferentially extracts lower-energy excitonic states. The observed longer lifetime further suggests the involvement of weakly allowed excitonic states, such as trion-related or dark-state transitions. While the exact microscopic origin remains to be clarified, the present observations are inconsistent with a simple Purcell-enhanced bright exciton emission mechanism.

\section{Summary}
In summary, we have demonstrated controllable photothermal effects and all-optical bandgap tuning in monolayer TMDC-integrated with an ultrahigh-Q silica microcavity. Quantitative analysis of the observed energy shifts of WSe$_2$ and WS$_2$ is well explained by the intracavity temperature rise induced by cavity-enhanced photothermal conversion. The photoluminescence emission collected through the fiber waveguide exhibits distinct spectral and temporal characteristics compared to free-space emission. These observations indicate that the high-Q microcavity selectively enhances specific emission channels, likely associated with long-lived excitonic states. Among various 2D material-integrated nanophotonic platforms, our study establishes a versatile approach for quantitatively probing and controlling photothermal effects in such hybrid systems.



\section*{Acknowledgments}
Part of this work was supported by JSPS KAKENHI (JP24K17624, JP24H01202, JP25H02153); MEXT Quantum Leap Flagship Program (JPMXS0118067246); Keio University Program for the Advancement of Next Generation Research Projects. S. Fujii acknowledges support from the Inamori Foundation and the Nippon Sheet Glass Foundation for Materials Science and Engineering. R. Sugano and Y. Takahashi acknowledge support from JST SPRING (JPMJSP2123).
The authors thank Hiroo Suzuki (Okayama University) for fruitful discussions on sample fabrication, and Yuichiro K. Kato (RIKEN) for helpful discussions.

\section*{Disclosures}
The authors declare no conflicts of interest.

\section*{Data availability}
Data underlying the results presented in this paper are obtained from the authors upon reasonable request.

\noindent
See Supplement 1 for supporting content.

	\bibliography{TMD_microsphere}
\end{document}